\documentstyle{article}

\def\abstract#1{\vskip 7mm 
        \begin{center}{\large Abstract}\par \smallskip
                \begin{minipage}[c]{12cm}
                        \small #1
                \end{minipage}
        \end{center}
}
\def\title#1{\begin{center}{\Large\bf #1}\end{center}}
\def\author#1{\vskip 5mm \begin{center}{#1}\end{center}}
\def\address#1{\begin{center}{\it #1}\end{center}}
\makeatletter
\@ifundefined{lesssim}{}{}
\@ifundefined{gtrsim}{}{}
\def\vereq#1#2{\lower3pt\vbox{\baselineskip1.5pt \lineskip1.5pt
\ialign{$\m@th#1\hfill##\hfil$\crcr#2\crcr\sim\crcr}}}
\makeatother

\newcommand{\vx}{{\bf x}}
\begin{document}

\title{%
  Multi-centered Solutions with a (very special)
  Warped Compactification
\footnote{
A short talk presented by K.S. at JGRG~`99 (November 3), at Hiroshima University, Japan.
}
}
\author{%
  Mitsuru Ooho\footnote{E-mail:c5024@sty.sv.cc.yamaguchi-u.ac.jp}
}
\address{%
  Graduate School of Science and Engineering, Yamaguchi University, \\
  Yoshida, Yamaguchi 753--8512, Japan
}
\author{%
  Kiyoshi Shiraishi\footnote{E-mail:shiraish@sci.yamaguchi-u.ac.jp}
}
\address{
  Graduate School of Science and Engineering, Yamaguchi University, \\
  Yoshida, Yamaguchi 753--8512, Japan \\
  and \\
  Department of Physics, Yamaguchi University, \\ 
  Yoshida, Yamaguchi 753--8512, Japan
}
\abstract{
We find an exact solution for multi-black strings
in the brane world with warped compactification. 
}

%%%%%%%%%%%%%%%%%%
\section{Introduction}
%%%%%%%%%%%%%%%%%%

Recently it has been suggested by much work on unification theories that
we live in a brane with three-dimensional spatial extension.
Randall and Sundrum proposed a warped compactification scenaro
\cite{RS1}.
\footnote{See also
\cite{akama0,rubakov,holdom,maia,callan,visser,squire,hughes,gibbons,akama1} for
earlier attempts.} They assumed the five-dimensional spacetime and the presence
of a negative cosmological constant in the bulk space. Two branes with tension of
individual values realize the compactification of an extra dimension on
$S^1/Z_2$. The mass scale on a brane depends on the warp factor which
is determined by the location of the brane in the extra dimension.
Thus their model provides a novel solution to the hierarchy problem.

Extensions to many branes have been studied by many authors.
(see, for example,
\cite{kobe}.)
\footnote{
A single brane scenario\cite{RS2} has been also considered.
We will not mention this scenario in detail here.
}

The models including dilaton fields have been also studied
\cite{Keha}
\cite{Haly}.
In the case,
although the bulk space is not an anti-de Sitter space,
similar warped compactification can be obtained.

On the other hand,
the black hole solution on the brane was analyzed by Chamblin et al.
\cite{CHR}.
Many other types of localized objects in the brane world
may be considered, and we expect that physics on a brane
will be clarified by study of them.

In this talk,
we show an exact solution for  charged objects in the brane world.

%%%%%%%%%%%%%%%%%%
\section{Brane world}
%%%%%%%%%%%%%%%%%%

We consider a model including a dilaton field
governed by the following action:
\begin{eqnarray}
S&=&\int d^Dx\sqrt{-g}\left[R-\frac{4}{D-2}\left(\nabla\phi\right)^2-
2\Lambda e^{4b\phi/(D-2)}\right] \nonumber \\
&-&\sum_k\int_{brane_k}d^{D-1}x\sqrt{-g_{D-1}}2\sigma_k e^{2b\phi/(D-2)},
\end{eqnarray}
where $\Lambda$ is a cosmological constant and $\sigma_k$ is 
the tension of the $k$th brane.
The metric is assumed to be
\begin{equation}
ds^2=\Omega^2(\chi)\left[\eta_{\mu\nu}dx^{\mu}dx^{\nu}+d\chi^2\right],
\end{equation}
where $\mu, \nu=0, \cdots, D-2$.
We also assume that the configuration of the dilaton field takes the form:
\begin{equation}
e^{4b\phi/(D-2)}=\Omega^{-2b^2}.
\end{equation}
Under the assumptions, field equations derived from the previous action
can be reduced to be
\begin{eqnarray}
(D-2)(D-1-b^2)\left(\frac{\Omega'}{\Omega}\right)^2&=&
(-2\Lambda)\Omega^{2-2b^2}, \\
-\left(\frac{\Omega'}{\Omega}\right)'+(1-b^2)\left(\frac{\Omega'}{\Omega}\right)^2&=&
\frac{1}{D-2}\Omega^{1-b^2}\sum_k\sigma_k \delta(\chi-\chi_k),
\end{eqnarray}
where $\chi_k$ is a position of the $k$th brane.

For $b^2=\!\!\!\!\!/ 1$, one can find the following solution:
\begin{eqnarray}
\left(\Omega^{b^2-1}\right)'&=&\pm
(1-b^2)\sqrt{\frac{(-2\Lambda)}{(D-2)(D-1-b^2)}}, \\
\left(\Omega^{b^2-1}\right)''&=&
(1-b^2)\frac{1}{D-2}\sum_k\sigma_k \delta(\chi-\chi_k).
\end{eqnarray}
This yields the brane solutions on tuning values for brane tensions.
\footnote{
For a single negative cosmological constant, one should choose
$\sigma_k=\pm 2\sqrt{-2\Lambda}$.
}

In particular, the case with $b^2=0$ corresponds to the Randall-Sundrum model.
To see this more explicitly, we introduce a new coodinate
$dz\equiv\Omega d\chi$.
Then the metric takes the form:
\begin{equation}
ds^2=e^{-2Kz}\eta_{\mu\nu}dx^{\mu}dx^{\nu}+dz^2,
\end{equation}
where $K$ is a certain constant.

For $b^2=1$, the solution is
\begin{equation}
\phi=\pm\frac{\sqrt{-2\Lambda}}{2}\chi,~~~\Omega=e^{-2\phi/(D-2)}.
\end{equation}
In this case with $b^2=1$,
we can simplifies the action, if we set
$\bar{g}_{\mu\nu}=e^{4\phi/(D-2)}g_{\mu\nu}$,
as
\begin{eqnarray}
\bar{S}&=&\int d^Dx\sqrt{-\bar{g}} e^{-2\phi}
\left[\bar{R}+4\left(\bar{\nabla}\phi\right)^2-
2\Lambda\right] \nonumber \\
&-&\sum_k\int_{brane_k}d^{D-1}x\sqrt{-\bar{g}_{D-1}}2\sigma_k e^{-2\phi}
\end{eqnarray}
Therefore, the bulk solution
for $b^2=1$ is the linear dilaton solution.

%%%%%%%%%%%%%%%%%%
\section{Black strings in the brane world}
%%%%%%%%%%%%%%%%%%

{}From now on, we concentrate our attention to the
$b=1$ case.
We introduce an anti-symmetric tensor field into the model:
\begin{eqnarray}
S&=&\int d^Dx\sqrt{-g}\left[R-\frac{4}{D-2}\left(\nabla\phi\right)^2-
\frac{1}{12}e^{-8\phi/(D-2)}H^2-
2\Lambda e^{4\phi/(D-2)}\right] \nonumber \\
&-&\sum_k\int_{brane_k}d^{D-1}x\sqrt{-g_{D-1}}2\sigma_k e^{2\phi/(D-2)},
\end{eqnarray}
where $H_{MNL}$ is an anti-symmetric tensor field strength.

In this model, a solution in the bulk space is:
\begin{equation}
ds^2=\Omega^2(\chi)\left[h^{-\frac{D-4}{D-2}}(-dt^2+d\chi^2)+
h^{\frac{2}{D-2}}\delta_{ij}dx^{i}dx^{j}\right],
\end{equation}
\begin{equation}
e^{4b\phi/(D-2)}=\Omega^{-2}h^{-\frac{2}{D-2}},~~~H_{t\chi i}=\partial_ih^{-1},
\end{equation}
\begin{eqnarray}
(D-2)^2\left(\frac{\Omega'}{\Omega}\right)^2&=&(-2\Lambda), \\
-\left(\frac{\Omega'}{\Omega}\right)'&=&
\frac{1}{D-2}\sum_k\sigma_k \delta(\chi-\chi_k),
\end{eqnarray}
with
\begin{equation}
h=1+\frac{1}{\Omega^{D-2}}\sum_a\frac{\mu_a}{(D-4)|\vx-\vx_a|^{D-4}},
\end{equation}
where $\mu_a$ are arbitrary constants.

A spacetime slice with a constant $\chi$ appears to be
extreme black holes with a dilaton and a scalar field:
\begin{equation}
ds^2_c=h^{\frac{D-4}{(D-2)(D-3)}}\left[-h^{-\frac{D-4}{D-3}}dt^2+
h^{\frac{1}{D-3}}\delta_{ij}dx^idx^j\right]+
h^{-\frac{D-4}{D-2}}d\chi^2.
\end{equation}
In particular, for $D=5$, the metric seems
\begin{equation}
ds^2_c=h^{\frac{1}{6}}\left[-h^{-\frac{1}{2}}dt^2+
h^{\frac{1}{2}}\delta_{ij}dx^idx^j\right]+
h^{-\frac{1}{3}}d\chi^2.
\end{equation}

%%%%%%%%%%%%%%%%%%
\section{Induced charge, energy density and pressure}
%%%%%%%%%%%%%%%%%%

When the black strings penetrate a brane perpendicularly,
matter and dilatonic charge distributions are ``induced'' on the brane.
For simplicity, we consider a single brane located at $\chi=0$.
The field equations are:
\begin{eqnarray}
& &R_{\mu\nu}-\frac{1}{2}Rg_{\mu\nu}-
\frac{4}{D-2}\left[\nabla_{\mu}\phi\nabla_{\nu}\phi-
\frac{1}{2}(\nabla\phi)^2g_{\mu\nu}\right] \nonumber \\
&-&\frac{1}{4}e^{-8\phi/(D-2)}\left[H^2_{\mu\nu}-\frac{1}{6}H^2g_{\mu\nu}
\right]+
e^{4\phi/(D-2)}\Lambda g_{\mu\nu}+
\frac{\sqrt{-g_{D-1}}}{\sqrt{-g}}e^{2\phi/(D-2)}\sigma
g_{\mu\nu}\delta(\chi) \nonumber \\
&=&\frac{4}{D-2}\frac{\sqrt{-g_{D-1}}}{\sqrt{-g}}e^{2\phi/(D-2)} T_{\mu\nu},
\end{eqnarray}
\begin{eqnarray}
&-&\frac{8}{D-2}\nabla^{2}\phi-
\frac{1}{12}\frac{8}{D-2}e^{-8\phi/(D-2)}H^2 \nonumber \\
&+&\frac{8}{D-2}e^{4\phi/(D-2)}\Lambda+
\frac{4}{D-2}\frac{\sqrt{-g_{D-1}}}{\sqrt{-g}}e^{2\phi/(D-2)}\sigma\delta(\chi)
\nonumber \\
&=&\frac{4}{D-2}\frac{\sqrt{-g_{D-1}}}{\sqrt{-g}}e^{2\phi/(D-2)}Q,
\end{eqnarray}
where $T_{\mu\nu}$ represents the ``induced'' energy-momentum tensor
and $Q$ represents the ``induced'' dilatonic charge on the brane,
when the bulk solution of black strings in the previous section is substituted
into the equations.

We obtain:
\begin{eqnarray}
\rho&=&-\sigma\left(1-\frac{1}{\sqrt{h}}\right)\delta(\chi), \\
p&=&-\frac{\sigma}{2}\left(\sqrt{h}+\frac{1}{\sqrt{h}}-2\right)
\delta(\chi), \\
Q&=&-\rho,
\end{eqnarray}
where
$T^{\mu}_{\nu}=diag.(-\rho,p,\cdots,p)$.
Here we omitted the singularity of the string core.

In the case with a single black string at the origin in $D=5$,
$\rho$ behaves as $1/r$ and $p$ as $1/r^2$, asymptotically
at a large distance $r$.

%%%%%%%%%%%%%%%%%%
\section{Summary}
%%%%%%%%%%%%%%%%%%

We have obtained an exact solution for multi-black strings
in the brane world with warped compactification. 
Since the black strings ``induce'' matter and charge distributions on the brane,
we suppose that the theory of gravity on the brane shoud be an
``unusual'' one.

In future work, we will examine the geodesic motion around the black strings
in the brane world.
We will also study black hole solutions with and without charges
in the models including dilatonic fields, and in the case with a general number
of extra dimensions.
It is also interesting to study black holes in the case with thick domain walls.

Furthermore, we should investigate the possible relation to the string/M theory.

%%%%%%%%%%%%%%%%%%%%%%%%%%%

\bigskip

\noindent
{\bf Note added}:
After this talk, we found \cite{Youm}, in which various solitons
are studied in the brane world with non-trivial dilatons. 
The authors would like to thank Prof. K.~Akama for information on the earlier
work concerning warped compactifications.

%%%%%%%%%%%%%%%%%%%%%%%%%%%

\end{document}